\documentstyle[a4,12pt,epsf]{article}

\begin{document}

\title{
{\bf Nuclear multifragmentation induced by  
electromagnetic fields of ultrarelativistic heavy ions} }
    
\author{I.A. Pshenichnov$^{1,2}\dagger$, I.N. Mishustin$^{2,3}\ddagger$, \\
J.P. Bondorf$^2$, A.S. Botvina$^{1,4}$, A.S. Iljinov$^1$\\ 
{\em $^1$ Institute for Nuclear Research, Russian Academy of Science,}\\
{\em 117312 Moscow, Russia}\\
{\em $^2$ Niels Bohr Institute, DK-2100 Copenhagen, Denmark}\\
{\em $^3$ Kurchatov Institute, Russian Research Center,}\\
{\em 123182 Moscow, Russia}\\
{\em $^4$ Hahn-Meitner-Institut, D-14109 Berlin, Germany}\\
}
\date{ }
\maketitle

\begin{abstract}
We study the disintegration of nuclei by strong electromagnetic fields
induced by ultrarelativistic heavy ions. The proposed multi-step model
includes 1) the absorption of a virtual photon by a nucleus, 2) intranuclear 
cascades of produced hadrons and 3) statistical
decay of the excited residual nucleus. The combined model describes well 
existing data on projectile fragmentation at energy  200 GeV per
nucleon. Electromagnetic multifragmentation of nuclei is predicted to be 
an important reaction mechanism at RHIC and LHC energies.  
\end{abstract}


$^\dagger$ E-mail: pshenichnov@al10.inr.troitsk.ru

$^\ddagger$ E-mail: mishustin@nbi.dk

\newpage

\section{Introduction}

Many experiments with 
relativistic heavy ions (RHI) are carried out now as well as planned for
the future. In these experiments the main interest is focused on 
nucleus-nucleus collisions with small impact parameters corresponding to a
significant overlap of nuclei. Central collisions are most appropriate for
studying the highly excited and compressed hadronic matter and the   
quark-gluon plasma. Peripheral collisions are used to study the 
fragmentation of spectators. 

However, the collisions with impact parameters exceeding the
sum of nuclear radii, i.e. with no direct overlap of nuclei,
can also significantly contribute to the nuclear reactions 
via the long--range electromagnetic 
interaction. The relativistic Coulomb excitation of nuclei was 
studied by many authors, see e.g. refs.~\cite{Winther,BertBaur,Vasconcellos}. 
This process is especially important in ultrarelativistic 
collisions with the Lorentz factors $\gamma \gg 1$. 
During a short time when colliding nuclei are close to each other, the  
potential of Lorentz--boosted Coulomb field~\cite{JDJackson} is very strong, 
$V_c \sim \alpha \gamma Z/b$, where $Z$ is the nuclear charge and 
$b$ is an impact parameter, $\alpha$ is the fine structure constant. 
For typical values $\gamma \gg \alpha^{-1}$, $Z\sim 50$
and $b \sim 10$ fm this 
potential $V_c \sim \alpha \gamma$~GeV considerably exceeds the nuclear 
forces and one can expect very dramatic phenomena. 
Among other interesting effects such an electromagnetic field will lead to 
the complete disintegration of 
nuclei, a phenomenon known already in nuclear reactions under the name of 
"multifragmentation"~\cite{JPB,Gross}.
The main aim of the present paper is to estimate the importance of these
electromagnetic processes and to draw attention 
to a potentially rich area of research associated with the behaviour of 
nuclear matter under very strong electromagnetic forces.

As a first step we describe these reactions within a simplified approach 
known as the Weizs\"{a}cker-Williams method. According to this method
one can represent the Lorentz--boosted Coulomb field 
as a pulse of radiation of virtual photons~\cite{BertBaur,JDJackson}.
Within this approach we use the virtual photon spectrum  calculated for
the point-like charge distributions, that is justified only for collisions 
without direct nuclear overlap. As it was found in~\cite{BaronBaur}, the
correction of the spectrum due to the finite
sizes of colliding nuclei is of minor importance for the total fragmentation 
cross section.

The aim of the present paper is threefold. Firstly, we present a model of
electromagnetic interaction of relativistic heavy ions which 
makes it possible to calculate the
inclusive and exclusive characteristics of such interactions and compare them 
with existing experimental data. Secondly, we calculate the 
electromagnetic dissociation (ED) cross sections and the yields  
of hadrons and nuclear fragments produced due to the
electromagnetic interaction of ultrarelativistic heavy ions which
will be available at RHIC and LHC colliders.
Thirdly, in order to evaluate the probability of formation
of very hot nuclei, 
we calculate the distributions in excitation energies of the 
residual nuclei formed after completion of the cascade stage of the reaction.

\section{Electromagnetic dissociation of nuclei}

\subsection{Preliminary remarks}

The process of electromagnetic dissociation  is known for a long time. 
Since only relatively low beam energies were available, mainly the 
nuclear evaporation and fission processes 
were studied up to now. In refs.~\cite{CBR}-\cite{GSingh2} the 
interaction of low-Z ions ($^{16}{\rm O}, ^{28}{\rm Si}, ^{32}{\rm S}$) 
with energies 
0.7--200A GeV was investigated via the projectile fragmentation in nuclear 
emulsions or on different targets (${\rm H, C, Al, Cu, Ag, Sn, Pb}$).
The electromagnetic dissociation of target nuclei has been also studied
experimentally~\cite{MTMercier}-\cite{JCHill3}, mainly via the measurement of 
$\gamma$-activity of target fragments. 

The electromagnetic interaction of high-Z ultrarelativistic heavy ions 
(with Lo\-rentz factors $\gamma \gg 1$) has attracted recently a special 
attention  in connection with the storage of the relativistic heavy-ion
beams. It turned out that the cross section of electromagnetic dissociation
of nuclei, 
$\sigma_{ED}$ becomes comparable with or even larger
than the total nuclear cross section. The first observation of $\sigma_{ED}$
values above 1 barn was reported for the interaction of $^{139}{\rm La}$ 
projectiles with a $^{197}{\rm Au}$ target~\cite{JCHill1}. 

There are some estimates for the cross 
sections expected for the planned relativistic 
heavy-ion colliders: RHIC at Brookhaven ($\sqrt{s}=200A$ GeV) 
and LHC at CERN ($\sqrt{s}=6.5A$ TeV). 
As shown by Baur and Bertulani~\cite{Baur1} for $^{238}{\rm U}$ on 
$^{238}{\rm U}$ and 
Mercier et al.~\cite{MTMercier} for $^{197}{\rm Au}$ on $^{197}{\rm Au}$, 
at RHIC energy $\sigma_{ED}$ 
reach values of 40 and 23.8 b, respectively, with the corresponding nuclear 
cross sections  of 6.9 and 6.1 b. This means that the lifetime of stored 
beams will be 
greatly affected by the ED process. Also the particle production in ED 
will represent a significant background in experiments on future heavy-ion 
colliders such as RHIC and LHC. 

The nature of a photonuclear interaction depends on the 
de Broglie wavelength $\lambda \hspace{-0.2cm}^{-}=\hbar /p$ 
of a virtual photon.
At $E_{\gamma}\leq$40 MeV, $\lambda \hspace{-0.2cm}^-$ is 
comparable with the nuclear size and an excitation of the whole 
nucleus in the form of a Giant Dipole Resonance (GDR) is most important
in this region.  
In the energy range 40$\leq E_\gamma \leq 140$ MeV 
(below the pion production threshold),
$\lambda \hspace{-0.2cm}^-$
is comparable to the internucleonic distance, and therefore the 
absorption of a photon by a neutron-proton pair should be the main 
reaction mechanism.
At higher energies, when $\lambda \hspace{-0.2cm}^-$  becomes 
smaller than the nucleon radius, photons interact mainly with
individual nucleons. This process is known as the photoproduction of 
hadrons. It is followed
by the emission of fast nucleons and pions from the nucleus.
In present work the photoabsorption process on a nucleus is considered 
in the framework of the IntraNuclear Cascade model (INC) for photonuclear 
reactions~\cite{Iljinov}. 

Although soft photons (i.e. with energy $E_{\gamma }\leq $100 MeV) dominate 
in the virtual photon spectrum, it extends up to several GeV 
for RHI with $\gamma \sim 10-100$.  
Most theoretical works on electromagnetic 
disintegration~\cite{Baur1}-\cite{WJLlope} study 
the excitation of GDR through the absorption of one or more virtual photons 
prior to the de-excitation stage.
Less efforts have been made to understand the nature of ED induced by the  
virtual photons from the high-energy part of the spectrum. 
We know the only calculation of inclusive pion production cross 
section~\cite{CABert} and the only experimental observation of pions 
in ED~\cite{GSingh2}, unfortunately with a very poor statistics.
In the case of multiple pion photoproduction the nucleus can absorb a large
amount of energy, that, in its turn, may result not only in the 
emission of one or two
nucleons or fission, but also in a multifragment break-up of the nucleus.

\subsection{Description of the model}

Since a relativistic particle spends a short time $\Delta t$
near the collision partner, the virtual photon spectrum will contain all
frequencies up to a maximum one of the order $\omega_{max}\sim 1/\Delta t$.
The magnitude of the pulse is proportional to the square of the charge.
Both maximum frequency and magnitude depend on the value of
impact parameter $b$~\cite{JDJackson}. 

According to the Weizs\"{a}cker-Williams method  
the spectrum of virtual photons from a stationary target of charge $Z_t$ 
as seen by a projectile moving with velocity $\beta =v/c$ is 
obtained by the integration over impact parameters $b > b_{min}$  
and is expressed as~\cite{JDJackson}:
\begin{equation}
n(E_\gamma)=\frac{2\alpha Z^{2}_t}{\pi \beta ^{2}E_\gamma}\Bigl( 
xK_0(x)K_1(x)-\frac{\beta^{2}}{2}x^{2}(K^{2}_{1}(x)-K^{2}_{0}(x))\Bigl). 
\end{equation} 
Where $K_0$ and $K_1$ are the modified Bessel functions of zero and first 
order, $x=E_\gamma b_{min}/(\gamma \beta \hbar c)$,  $b_{min}$ is the 
minimal value of the impact parameter which corresponds to the onset of nuclear
interaction. With the curvature correction this
value can be estimated as
\begin{equation}
b_{min}=r_0\Bigl(A^{1/3}_p+A^{1/3}_t-X(A^{-1/3}_p+A^{-1/3}_t)\Bigr), 
\end{equation}
where $r_0=$1.34~fm, $A_p$ and $A_t$ are the nuclear numbers in 
projectile and target nuclei, respectively, and the curvature parameter 
$X=$0.75~\cite{MTMercier}. 
The cross section of ED for the projectile 
of mass $A_p$ can be calculated as:
\begin{equation}
\sigma_{ED}=\int\limits_{0}^{\infty}n(E_\gamma)\sigma_{A_p}(E_\gamma)dE_\gamma
\label{eq:3}
\end{equation}
where $\sigma_{A_p}(E_\gamma)$ is the appropriate photoabsorption cross 
section measured for the projectile nucleus with real photons~\cite{JAhrens} 
or calculated by a model.

A realistic estimation of $\sigma_{ED}$ can be done as follows.
In the GDR and quasideuteron regions, $E_\gamma\leq$ 140 MeV, one can utilize
the tabulated values of $\sigma_{A_p}(E_\gamma)$. Above the pion production 
threshold a universal behaviour of $\sigma_{A_p}(E_\gamma)$ can be used 
since the ratio $\sigma_{A_p}(E_\gamma)/A_p$ depends weakly  on $A_p$ 
up to $E_\gamma =2-3$ GeV. Having the data for one nucleus one can 
calculate the 
cross section for other nuclei~\cite{NBianchi}. 
In this region, the excitation of numerous
baryonic resonances (mainly $\Delta (1232), N^\star (1520), N^\star (1680)$)
provides a prominent variation of the photoabsorption cross section,
even though most of the resonances suffer damping inside the 
nucleus~\cite{LAKondratyuk}.
Above $E_\gamma\sim 3$~GeV the ratio  
$\sigma_{A_p}(E_\gamma)/A_p$ becomes lower for heavy nuclei due to 
the well-known shadowing effect~\cite{JAhrens}. This effect, which might 
be explained by
the hadronic component of the photon wave function, is not very strong,
of the order of 20\% , even for heaviest nuclei.    

\begin{figure}[hp]  
\begin{centering}
\epsfxsize=0.9\textwidth
\epsffile{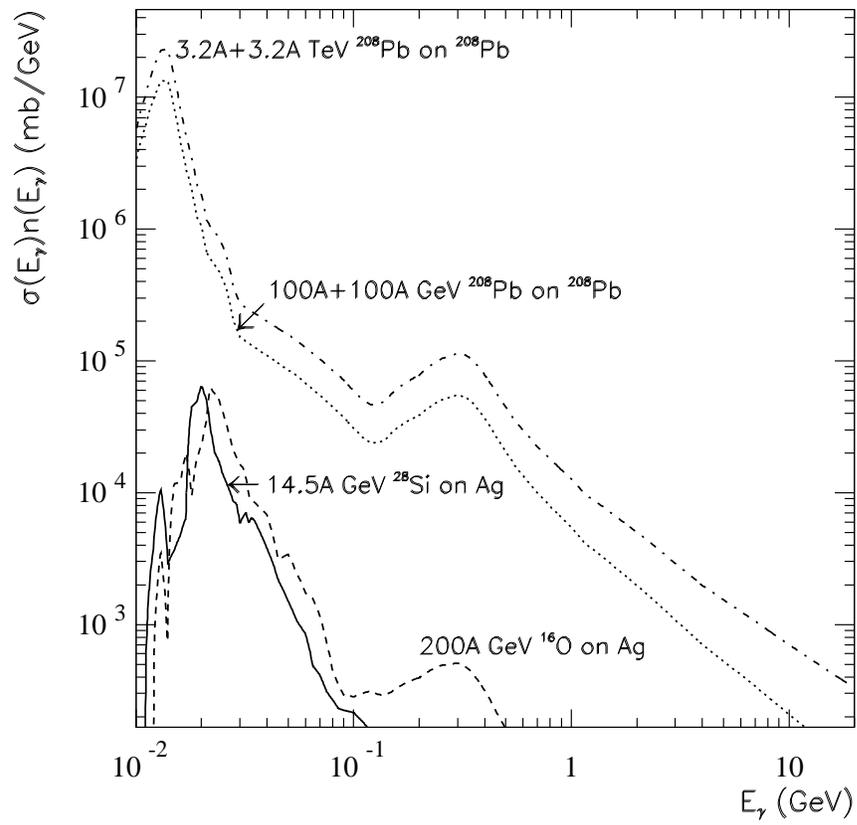}
\caption{
Product $n(E_\gamma)\sigma_{A_p}(E_\gamma)$ as a function of
virtual photon energy $E_\gamma$.
}
\label{fig:1}
\end{centering}
\end{figure}

Using the Monte Carlo technique, we have first simulated the energy $E_\gamma$
of a virtual photon distributed according to the product 
$n(E_\gamma)\sigma_{A_p}(E_\gamma)$, which 
is proportional to the probability of the electromagnetic interaction.  
This function is plotted in Fig.~\ref{fig:1} for the cases of RHI 
collisions considered in the paper.  
As seen from Fig.~\ref{fig:1},
the probability to have a photonuclear reaction above the pion production
threshold is essentially enhanced for RHI beams with $\gamma \gg $100. 
The second step is the nuclear photoabsorption process which is
treated on the basis of the IntraNuclear Cascade model (INC).

In ref.~\cite{JPB} we have already shown that the INC model can be used
for the description of reactions leading to the nuclear  
multifragmentation. This approach can be easily generalized for the 
photonuclear processes.
At high photon energies many hadrons 
are produced inside a nucleus in the course of a cascade process. 
Due to the knock-out of the intranuclear nucleons by 
fast nucleons and pions, the slower particles pass through a lower density 
region undergoing less rescatterings, the 
so-called trawling effect. In the present calculations in 
addition to the standard version of the INC model we
use also the version with trawling, which takes into account the local
depletion of nuclear density during the development of the intranuclear 
cascade. 
As was shown in~\cite{bot90,golu94}, the trawling effect is important for 
a realistic description of reactions at projectile energies above several GeV.
The annihilation of energetic antiprotons is an example of the processes where
heating of nuclei by a multi-pion system takes place.
Recently it was found~\cite{FGoldenbaum} that the distributions of 
excitation energies of residual nuclei produced in this process
are in a good agreement with 
predictions of the INC model with trawling.
In the present paper the calculations with the two INC versions can be 
considered as a measure of uncertainty in describing the cascade stage.

The model of photonuclear reactions~\cite{Iljinov} used below takes into 
account qua\-si\-deute\-ron absorption of virtual photons as well as the hadron 
production on intranuclear nucleons. We disregard 
the multiple excitation processes with the absorption of more than one 
photon prior to the de-excitation stage. As shown by
Llope and Braun-Munzinger~\cite{WJLlope}, the single photon absorption 
dominates at beam energies above $\sim 10$A~GeV. Our estimates 
analogous to those made in ref.~\cite{BertBaur} show
that the probabilities for absorbing simultaneously 
two or three photons with
energies above 0.4 GeV are less than 9\% and 1.6\% , respectively, even
for ${\rm Pb}+{\rm Pb}$ collisions at LHC energies.  
  
The channels of the hadron photoproduction on a nucleon 
which are taken into account in the model are listed in Table~\ref{T1}. 
The two-body channel $\gamma N \rightarrow \pi N$ dominates up to
 $E_\gamma \sim$ 0.5 GeV, while  at energies $0.5 \leq  E_\gamma \leq 2$ GeV
the channels $\gamma N \rightarrow 2 \pi N$ and
$\gamma N \rightarrow 3 \pi N$ play a major role. However in this 
energy region the contribution of two-body
subchannels $\gamma N \rightarrow \pi B^\star$ and
$\gamma N \rightarrow M^\star N$ ($B^\star$ and $M^\star$ are baryon and
 meson resonances, respectively) is also prominent. Finally, when the photon
 energy reaches several GeV, the multiple pion production
becomes most important. A large number ($\sim 80$) of many-body
subchannels contribute to this process. As shown in ref.~\cite{Iljinov}, 
the INC model describes well the available
data on the meson and proton production  on nuclei by quasi-monochromatic
photons up to energies $E_{\gamma} \sim 10$GeV, that covers the main 
part of the total cross-section of ED reactions.

Excited residual nuclei are formed after the completion of
the  intranuclear cascades. The ensemble of residual nuclei is characterized
by wide distributions in excitation
energy,  $E^\star$, nucleon, $A$, and proton, $Z$, numbers.
The decay of the residual nuclei is described by the Statistical 
Multifragmentation Model (SMM)~\cite{JPB}. 
The SMM assumes that a hot nucleus expands  to a "freeze-out" 
volume where it splits into primary hot fragments and nucleons in thermal 
equilibrium. The break-up channels are constrained by the total mass, charge 
and energy of the system. All fragments (and nucleons) are considered as 
Boltzmann particles while Fermi-gas 
approximation is used for their internal excitation. 
The probabilities of different break-up channels are calculated 
microcanonically according to their statistical weights. After 
primary break-up excited fragments propagate independently under 
mutual Coulomb field and undergo secondary decay described by 
evaporation, Fermi-break-up or fission depending on their mass 
and excitation energy. The simulation of the whole process 
is  performed by the Monte Carlo method.

The main conclusions of the SMM are as follows.
The decay of a residual nucleus is determined by its  excitation
energy per nucleon $E^\star /A$.  For $E^\star /A < 2$ MeV, the
de-excitation goes through the successive emission of
particles from the compound nucleus (evaporation), or its fission. When the 
excitation energy of the nucleus exceeds about a half its total binding energy 
($E^\star /A > 4$  MeV), the explosive multifragment break-up  
dominates. In the transition region  
(2$\leq E^\star /A\leq$ 4 MeV) the both decay processes coexist.
At $E^\star /A > 10$ MeV the system breaks up into nucleons and
lightest clusters (vaporization).

This hybrid approach combining the INC and SMM models is quite successful 
in explaining the  fragmentation and multifragmentation data (see examples 
in refs.~\cite{JPB,bot90}). However, as was pointed out in ref.~\cite{bot92}, 
there are signs of early particle emission which accompany the 
after-cascade evolution of the residual nuclei towards the point of 
fragment separation at freeze-out. Therefore thermal excitation energies 
and masses of thermalized systems at multifragmentation could be smaller than 
calculated with the INC model. From our previous experience~\cite{JPB}
we can approximately estimate an uncertainty introduced by this effect as 
no more than factor 2 in inclusive yields of fragments.

\subsection{Analysis of existing data}

\begin{figure}[hp]  
\begin{centering}
\epsfxsize=0.9\textwidth
\epsffile{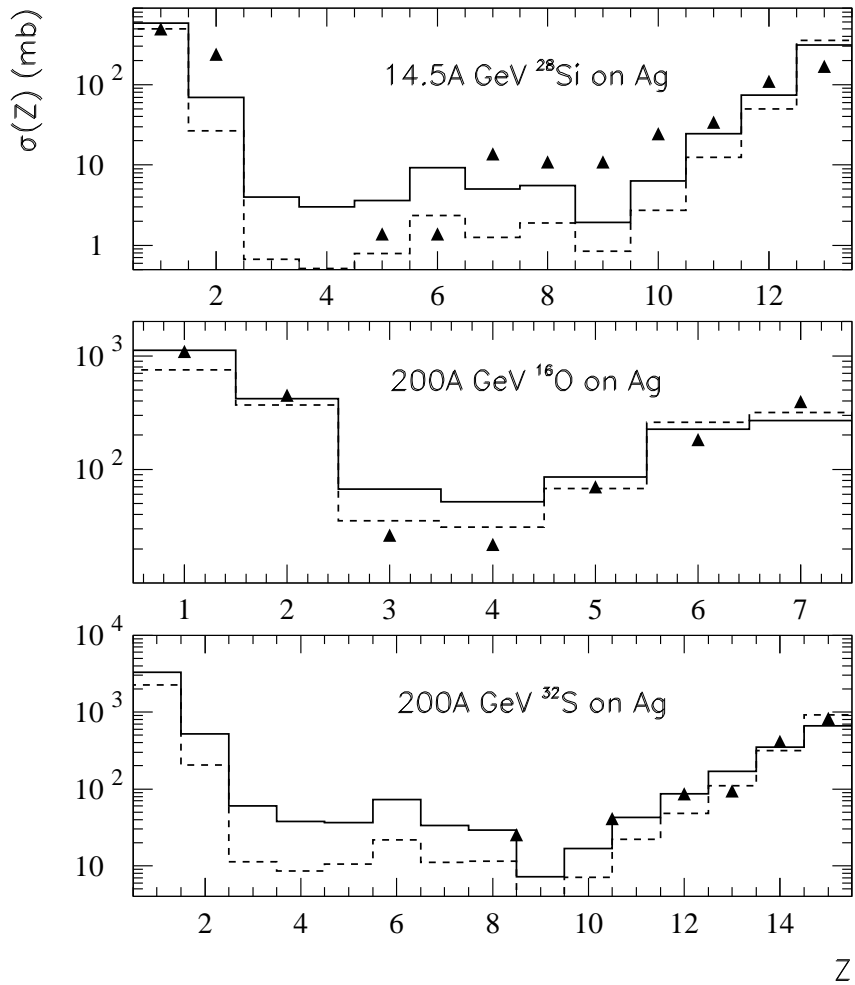}
\caption{
Inclusive cross sections of different isotope production 
in the electromagnetic dissociation of the projectile as a function of 
isotope charge $Z$. Triangles are the experimental 
data \protect\cite{GBA,GSingh2}. 
Calculation results for two versions of the INC model, with and without 
trawling, are presented by dashed and solid histograms, respectively. 
}
\label{fig:2}
\end{centering}
\end{figure}

With this combined approach we are able to calculate the yields of different 
isotopes in the ED process and compare them with existing experimental data. 
The isotope yields were measured in several 
experiments~\cite{GBA,GSingh2}. As it is seen from Fig.~\ref{fig:2}, 
a wide range of isotopes is produced in the disintegration of light nuclei.
Both versions of the INC model show a satisfactory agreement for 200A GeV 
ions, but for lower ion energy (14.5A GeV) the agreement is poor.  
Since our model is designed specially for RHI with $\gamma \gg 10$, we 
consider these results as acceptable.

In the emulsion experiments~\cite{GBA,GSingh2}
each decay mode into charged particles can
be measured, and its fraction to the total number of observed projectile
dissociation events can be determined. 
The calculated fraction of events in different groups of disintegration 
channels is given in Table~\ref{T2}. 
Taking into account that different experiments give different values, we 
conclude that the model describes these data fairly well. 

\begin{figure}[hp]  
\begin{centering}
\epsfxsize=0.8\textwidth
\epsffile{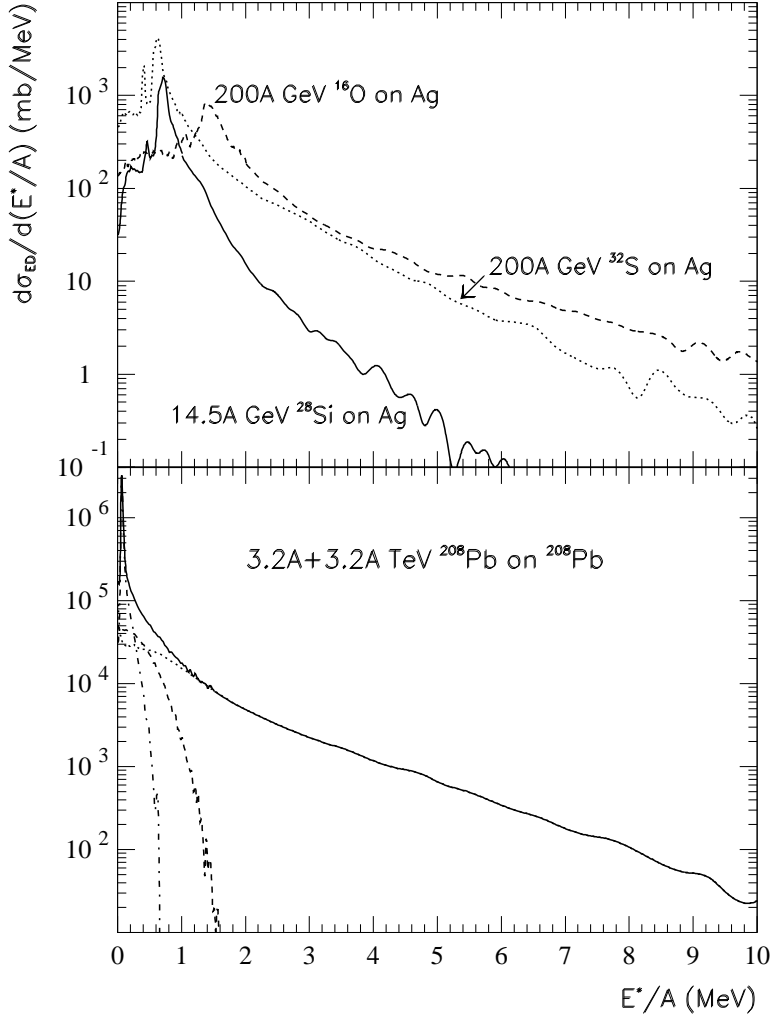}
\caption{
Distribution of excitation energies (per nucleon) of residual projectile 
nuclei produced in the electromagnetic interactions of the projectile 
and target nuclei. The calculations are presented for the INC model with 
trawling only. Results for 14.5A GeV $^{28}{\rm Si}$ on ${\rm Ag}$, 
200A GeV $^{16}{\rm O}$ and $^{32}{\rm S}$ on ${\rm Ag}$ are shown 
respectively  
by the solid, dashed and dotted lines in the top panel.
Predictions for $^{208}{\rm Pb}+^{208}{\rm Pb}$ collisions at LHC energy  
are shown by the solid line in the bottom panel. 
Contributions from the processes of GDR, $\Delta$-isobar excitation and 
multiple pion production are shown by the dot-dashed,
dashed and dotted lines, respectively.}
\label{fig:3}
\end{centering}
\end{figure}

As already mentioned, the decay mechanism of a residual nucleus
depends on its excitation energy per nucleon. It is seen from 
Fig.~\ref{fig:3} that the distribution of the excitation energies is very broad 
and tightly connected with the mass number of the initial nucleus and the 
projectile energy. The lighter nucleus ($^{16}{\rm O}$ in comparison with 
$^{32}{\rm S}$)
receives higher excitation energy per nucleon.    
For all reactions shown in Fig.~\ref{fig:3} only 
a limited fraction of residual nuclei excited by virtual photons
receives an excitation energy per nucleon sufficient for 
multifragmentation.
However their absolute yields are large in comparison with those obtained in 
usual nuclear reaction. It is the multiple pion production process that is 
responsible for production of heavy nuclei with $E^\star /A$ exceeding 
2 MeV.

\section{Predictions for ultrarelativistic heavy-ion 
collisions}

As the above discussion shows, 
the most interesting ED processes should be expected in ultrarelativistic 
heavy-ions collisions at RHIC or LHC colliders~\cite{RHIC-LHC}.
The relative Lorentz factor $\gamma$ is about $2\cdot 10^4$ for RHIC 
and $2\cdot 10^7$ for LHC energies. 
The calculated cross sections $\sigma_{ED}$ for the ED process 
are given in Table~\ref{T3} together
with the nuclear fragmentation cross section $\sigma_{nuc}$ to be used for 
comparison.
It is assumed that $\sigma_{nuc}$ does not depend on the projectile energy
in the multi-GeV region and can be parametrized as~\cite{YDHe}:
\begin{equation}
\sigma_{nuc}=59(A^{1/3}_p+A^{1/3}_t-0.83)^2 (mb).
\end{equation}
The $\sigma_{ED}$ is obtained by expression~(\ref{eq:3}), where the
integration covers all energies of virtual photons.   
One can estimate the contribution, $\sigma_{ED}^\Delta$, to the ED cross 
section coming from the photoabsorption in the 
$\Delta$-resonance region by limiting integration to photon energies between 
0.14 and 0.4 GeV. The contribution from multiple pion production, 
$\sigma_{ED}^{M\pi}$, 
is obtained by integrating over photon energies above 0.4 GeV. The 
value $\sigma_{ED}^{IMF}$ is defined as a cross section
to produce Intermediate Mass Fragments (IMF) with charges specified in  
Table~\ref{T3}: $Z=3-5$, $3-10$ or $3-30$, depending on the mass of the 
projectile which undergoes the multifragmentation.

As seen from Table~\ref{T3}, the model  
slightly overestimates the experimental $\sigma_{ED}$ values.
One should bear in mind, however that $\sigma_{ED}$ is very
sensitive to the details of the total photoabsorption cross section,
$\sigma_{A_p}(E_\gamma)$, 
particularly in the GDR region. More accurate data for this cross section 
are certainly needed. The calculations show that the ED process becomes 
important already for heavy-ion collisions at CERN SpS 
(160A GeV $^{208}{\rm Pb}$ on  $^{208}{\rm Pb}$) which presently are under 
investigations.

Considering the values presented in the last two rows of Table~\ref{T3},
one can note that $\sigma_{ED}^{M\pi}$ exceeds $\sigma_{ED}^\Delta$ at
RHIC and LHC energies and these two processes are responsible for 30-40\% of 
the ED cross section.  This proves that the pion photoproduction processes 
must be taken into consideration. 
For ${\rm Pb}+{\rm Pb}$ case $\sigma_{ED}$ is more than 10 times
larger than $\sigma_{nuc}$. Even $\sigma_{ED}^{IMF}$ given by our  model, 
turns out to be comparable or even greater than $\sigma_{nuc}$, that makes 
it possible to measure the electromagnetically induced multifragmentation 
of ultrarelativistic heavy ions.

\begin{figure}[hp]  
\begin{centering}
\epsfxsize=0.9\textwidth
\epsffile{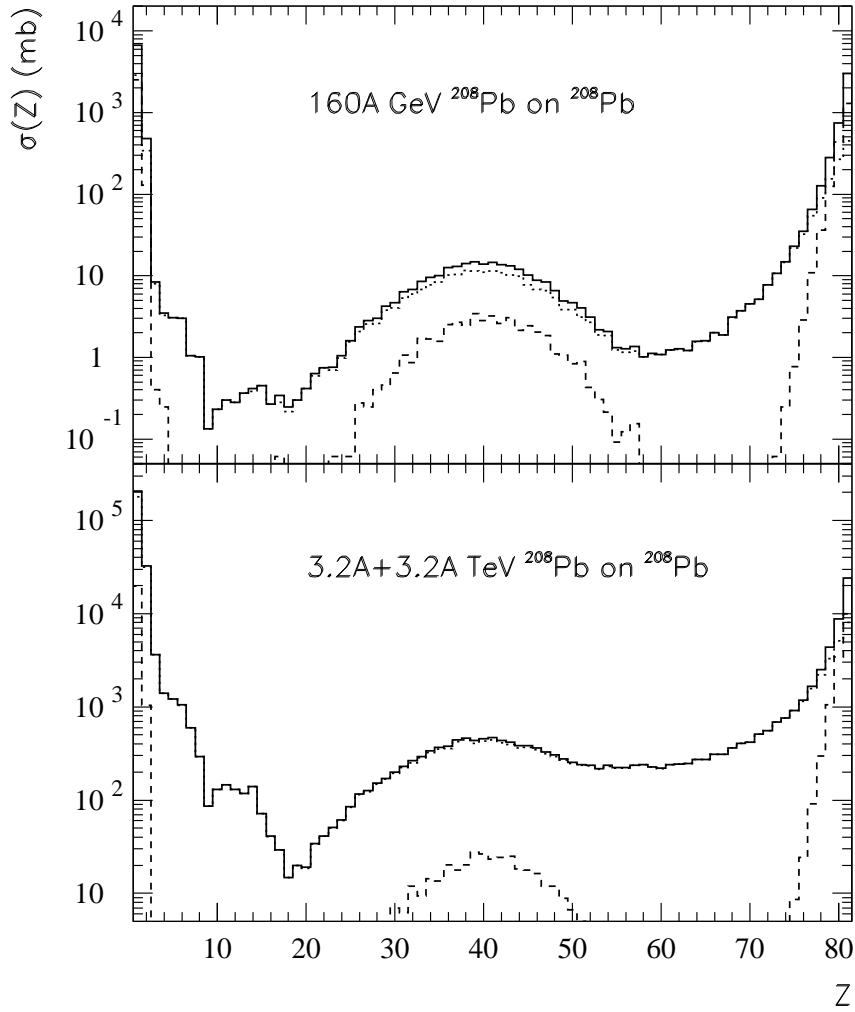}
\caption{
Inclusive cross sections of different isotope production 
in the electromagnetic dissociation of the projectile as a function of 
isotope charge $Z$. Predictions of the INC model with trawling are given.
Contributions from the processes of $\Delta$-isobar excitation and 
multiple pion production are shown by the
dashed and dotted histograms, respectively.   
}
\label{fig:4}
\end{centering}
\end{figure}

In Fig.~\ref{fig:4} we show predicted yields of fragments
produced in the ED process. It is seen that the total cross-sections 
are by one order of magnitude higher than the ones observed in traditional 
nuclear processes. 
At LHC energies (bottom part) $\sigma (Z)$-distribution is enriched by 
intermediate mass fragments.
The isotope yields are shown separately for singe and multiple pion 
production. 
Photons with energies in the $\Delta$-resonance region provide only a
moderate nuclear excitation. In this case the de-excitation goes mainly 
through the emission of several nucleons 
(the left and right parts of Fig.~\ref{fig:4}) or fission (the central part). 
The absorption of a high-energy photon leads to a multiple pion 
production and, consequently, to a higher excitation of a nucleus. 
This results in the deep disintegration of the nucleus and multiple 
production of nuclear fragments with a wide mass distribution.

The average pion multiplicity, $n_\pi$, calculated over the whole region
of virtual photon energies is not expected to be large, 
since the contribution from multiple pion photoproduction 
is not dominating in the ED process. Our calculations 
give $n_\pi\sim 1$ even for LHC energies, that is much lower 
than mean $n_\pi$ in nuclear collisions. 
But this is compensated by their large yields and broad multiplicity 
distributions. Inclusive cross
sections of proton emission and pion production in the ED process are given
in Table~\ref{T4}. As one can see, these cross sections 
are predicted to be up to ten times larger than the nuclear cross sections.

\section{Conclusions}

It is demonstrated that the electromagnetic dissociation of nuclei gives 
a dominant contribution to the total cross section in ultrarelativistic 
heavy ion collisions. The fragmentation cross-section 
for RHIC and LHC beams exceeds considerably the standard  nuclear cross 
sections. This means that the electromagnetic dissociation of 
ultrarelativistic beams on the residual gas should be considered seriously 
as a factor reducing the lifetime of the beams in RHIC or LHC 
colliders~\cite{RHIC-LHC}. 
Its contribution to the spectra and multiplicities 
of particles must be considered in the planned experiments. 

The electromagnetic fields generated by ultrarelativistic heavy ions
are so strong that one can expect qualitatively new phenomena in
these collisions. Nuclear multifragmentation is only one example  
of rich possibilities. 

The  long-range Coulomb forces  could lead to a new type of collective 
motion. For example, instead of radial flow of nuclear matter caused by 
its compression in central nucleus--nucleus collisions~\cite{JPB} 
one may expect a perpendicular flow caused by tidal Coulomb forces. 
The Weizs\"{a}cker-Williams method of virtual photons  might be not fully 
adequate to study these collective phenomena, especially at small impact 
parameters. We consider the present 
work only as a starting point for future studies.

We are grateful to J.J.Gaardh\o{j}e and L.M.Satarov for useful discussions. 
I.N.M is grateful to the Carlsberg Foundation for the financial support.
A.S.B and I.A.P. thank the Niels Bohr institute for the warm 
hospitality and financial support. The work was supported partially 
by the Danish Natural Science Research Council and INTAS, grant 93-1560.

\newpage

\begin{centering}
\begin{table}[ht]
\caption{Exclusive channels of the elementary $\gamma N$ interaction taken
 into account in the INC calculations.}
\vspace{0.3cm} 
\begin{tabular}{|c|c|} \hline
$\gamma p$-interaction & $\gamma n$-interaction  \\[3mm] \hline
$\gamma  p \rightarrow \pi^+ n$  
&$\gamma  n \rightarrow \pi^- p$  \\ 
$\gamma  p \rightarrow \pi^0 p$   
&$\gamma  n \rightarrow \pi^0 n$  \\ 
               &         \\
$\gamma  p \rightarrow \pi^-  \Delta^{++}$ 
&$\gamma  n \rightarrow \pi^-  \Delta^{+}$ \\
$\gamma  p \rightarrow \pi^0  \Delta^{+}$ 
&$\gamma  n \rightarrow \pi^0  \Delta^{0}$ \\
$\gamma  p \rightarrow \pi^+  \Delta^{0}$ 
&$\gamma  n \rightarrow \pi^+  \Delta^{-}$ \\
               &          \\
$\gamma  p \rightarrow \eta p$   
&$\gamma  n \rightarrow \eta n$  \\ 
$\gamma  p \rightarrow \omega p$   
&$\gamma  n \rightarrow \omega n$  \\ 
$\gamma  p \rightarrow \rho^0 p$   
&$\gamma  n \rightarrow \rho^0 n$  \\ 
$\gamma  p \rightarrow \rho^+ n$   
&$\gamma  n \rightarrow \rho^- p$  \\
               &            \\
$\gamma  p \rightarrow \pi^+  \pi^-  p$ 
&$\gamma  n \rightarrow \pi^+  \pi^- n $\\
$\gamma  p \rightarrow \pi^0  \pi^+  n$ 
&$\gamma  n \rightarrow \pi^0  \pi^- p $\\
              &              \\
$\gamma  p \rightarrow \pi^0  \pi^0   \pi^0  p$ 
&$\gamma  n \rightarrow \pi^0  \pi^0   \pi^0  n $\\
$\gamma  p \rightarrow \pi^+  \pi^-   \pi^0  p$ 
&$\gamma  n \rightarrow \pi^+  \pi^-   \pi^0  n $\\
$\gamma  p \rightarrow \pi^+  \pi^0   \pi^0  n$ 
&$\gamma  n \rightarrow \pi^-  \pi^0   \pi^0  p $\\
$\gamma  p \rightarrow \pi^+  \pi^+   \pi^-  n$ 
&$\gamma  n \rightarrow \pi^+  \pi^-   \pi^-  p $\\
                      &           \\
$\gamma  p \rightarrow i \pi  N (4 \leq i \leq 8)$
& $\gamma  n \rightarrow i \pi  N (4 \leq i \leq 8)$\\
(35 channels)    &  (35 channels) \\ \hline
\end{tabular}
\label{T1}
\end{table}
\end{centering}

\newpage

\begin{centering}
\begin{table}[ht]
\caption{Relative contribution of different decay modes for ED of 
the 200A GeV $^{16}{\rm O}$ projectile in emulsion.}
\vspace{0.3cm} 
\begin{tabular}{|l|c|c|c|c|}    
\hline\hline
Decay mode &\multicolumn{4}{c|}{Fraction (\% )} \\ \cline{2-5} 
\ & Standard INC & INC with trawling &\multicolumn{2}{c|}{Experiments}
 \\ \cline{4-5}
\ & and SMM & and SMM  & \cite{GBA} & \cite{GSingh2}  \\ 
\hline\hline  
$^{15}{\rm N}+p$   & 50.4 & 46.7 & 56.08$\pm $3.93   &   49.45$\pm $6.62   \\
\hline 
$^{12}{\rm C}+\alpha$  & 36.0 & 43.2 & 25.58$\pm $2.61 & 23.01$\pm $4.80   \\
$^{12}{\rm C}+2d$    &  \ &     &        \           &         \           \\ 
\hline
$^{11}{\rm B}+\alpha +p$  &  \   &  &    \           &         \           \\ 
$^{8}{\rm Be}+\alpha +2d$ &  8.6 & 6.26 & 4.42$\pm $1.10 & 10.62$\pm $3.06 \\ 
$^{8}{\rm Be}+^{7}{\rm Li}+p$   &  \   &  &    \     &         \           \\ 
\hline
$^{7}{\rm Li}+2\alpha +p$   &  \  &  &        \     &     \           \\ 
$^{7}{\rm Li}+\alpha +2d+p$ &  1.85 &0.95 & 2.49$\pm $0.83 & 4.42$\pm $1.98 \\ 
$^{7}{\rm Li}+4d+p$     &  \  &  &        \           &         \           \\ 
\hline
$       4\alpha   $      &  \  &  &     \           &         \           \\ 
$      3\alpha +2d$      &  2.9&3.0&8.01$\pm $1.49  &   12.39$\pm $3.31   \\ 
$2\alpha+4d$             &  \  &  &     \           &         \           \\ 
$ \alpha+6d$             &  \  &  &     \           &         \           \\ 
\hline\hline
\end{tabular}
\label{T2}
\end{table}
\end{centering}

\begin{centering}
\begin{table}[ht]
\caption{Cross sections of the ED process (barn) calculated by the 
model for the present and future RHI beams. See the text for 
notations.}
\vspace{0.3cm} 
\begin{tabular}{|c|c|c|c|c|c|c|c|}    
\hline\hline
  &  &\multicolumn{2}{c|}{\ } &  &  & \multicolumn{2}{c|}{\ } \\
Reaction & $\sigma_{nuc}$ & \multicolumn{2}{c|}{$\sigma_{ED}$} &
$\sigma_{ED}^\Delta$ & $\sigma_{ED}^{M\pi}$ 
&\multicolumn{2}{c|}{$\sigma_{ED}^{IMF}$} 
 \\ \cline{3-4}\cline{7-8} 
\ & \ & theory & experiment & \ & \ & standard & trawling \\
\hline\hline  
14.5A GeV    & 2.853 & 0.499& 0.383$\pm$0.037 & 0.020 & 0.002 &0.039&0.011  \\
$^{28}{\rm Si}\ on\ {\rm Ag}$ &       &      & \cite{GSingh2} &  & 
&\multicolumn{2}{c|}{$Z=3-10$} \\
\hline 
200A GeV     & 2.445 & 0.844& 0.592$\pm$0.057 & 0.110 & 0.080 &0.205&0.134 \\
$^{16}{\rm O}\  on\ {\rm Ag}$ &       &      & \cite{GSingh2} &  &    
&\multicolumn{2}{c|}{$Z=3-5$} \\
\hline 
200A GeV     & 2.968 & 1.643& 1.680$\pm$0.080 & 0.213 & 0.145 &0.294&0.086 \\
$^{32}{\rm S}\  on\ {\rm Ag}$ &       &      & \cite{GBA}     &  &  
&\multicolumn{2}{c|}{$Z=3-10$} \\
\hline 
160A GeV        & 7.164 & 40.00&    ---  &  3.070&1.662  &0.857 & 0.091  \\
$^{208}{\rm Pb}\ on \ ^{208}{\rm Pb}$ &  &   &                &  & 
&\multicolumn{2}{c|}{$Z=3-30$} \\
\hline
100A+100A GeV        & 7.164 &101.15&    ---  & 11.50 &16.74  & 30.42&5.12 \\
$^{208}{\rm Pb}\ on \ ^{208}{\rm Pb}$ &  &   &                &  & 
&\multicolumn{2}{c|}{$Z=3-30$} \\
\hline
3.2A+3.2A TeV        & 7.164 &202.7 &    ---  &  23.92 & 42.46 &79.51 & 15.0 \\
$^{208}{\rm Pb}\ on \ ^{208}{\rm Pb}$ &  &   &                &  & 
&\multicolumn{2}{c|}{$Z=3-30$} \\
\hline\hline
\end{tabular}
\label{T3}
\end{table}
\end{centering}

\newpage

\begin{centering}
\begin{table}[ht]
\caption{Inclusive cross sections (barn) of proton emission and pion 
production in the electromagnetic dissociation of the projectile 
calculated by the INC model with trawling for the present 
and future RHI beams.}
\vspace{0.3cm} 
\begin{tabular}{|c|c|c|c|c|c|}    
\hline\hline
  &  &\multicolumn{4}{c|}{\ } \\
Reaction & $\sigma_{nuc}$ & \multicolumn{4}{c|}{$\sigma_{ED}$} \\ 
\cline{3-6} 
\ & \ & $p$ & $\pi^+$ & $\pi^-$ & $\pi^0$ \\
\hline\hline  
14.5A GeV    & 2.853 & 0.492& 0.003 & 0.003 & 0.006  \\
$^{28}{\rm Si}\ on\ {\rm Ag}$ &       &      &  &  &  \\
\hline 
200A GeV     & 2.445 & 0.623& 0.059 & 0.062 & 0.078  \\
$^{16}{\rm O}\  on\ {\rm Ag}$ &       &      &  &  &  \\
\hline 
200A GeV     & 2.967 & 2.19 & 0.105 & 0.110 & 0.144  \\
$^{32}{\rm S}\  on\ {\rm Ag}$ &       &      &  &  &   \\
\hline
160A      GeV     & 7.164 &  9.72& 1.04    &  1.68 & 2.00   \\
$^{208}{\rm Pb}\ on \ ^{208}{\rm Pb}$ &   &      &  &  &   \\
\hline
100A+100A GeV     & 7.164 & 89.6 & 17.9    & 23.9  & 27.1   \\
$^{208}{\rm Pb}\ on \ ^{208}{\rm Pb}$ &   &      &  &  &   \\
\hline
3.2A+3.2A TeV     & 7.164 & 241.7& 52.7    &  68.9 &  77.4  \\
$^{208}{\rm Pb}\ on \ ^{208}{\rm Pb}$ &   &      &  &  &   \\
\hline\hline
\end{tabular}
\label{T4}
\end{table}
\end{centering}

\newpage


\end{document}